# Enhanced sensitivity of silicon-photonics-based ultrasound detection via BCB coating

**Resmi Ravi Kumar[1], Evgeny Hahamovich[1], Shai Tsesses[1], Yoav Hazan[1], Assaf Grinberg[1], Amir Rosenthal[1]**

[1]Andrew and Erna Viterbi Faculty of Electrical Engineering, Technion - Israel Institute of Technology

**Abstract:** Ultrasound detection via silicon waveguides relies on the ability of acoustic waves to modulate the effective refractive index of the guided modes. However, the low photo-elastic response of silicon and silica limits the sensitivity of conventional silicon-on-insulator (SOI) sensors, in which the silicon core is surrounded by a silica cladding. In this paper, we demonstrate that the sensitivity of silicon waveguides to ultrasound may be significantly enhanced by replacing the silica over-cladding with bisbenzocyclobutene (BCB) – a transparent polymer with a high photo-elastic coefficient. In our experimental study, the response to ultrasound, in terms of the induced modulation in the effective refractive index, achieved for a BCB-coated silicon waveguide with TM polarization was comparable to values previously reported for polymer waveguides and an order of magnitude higher than the response achieved by an optical fiber. In addition, in our study the susceptibility of the sensors to surface acoustic waves and reverberations was reduced for both TE and TM modes when the BCB over-cladding was used.

## 1. Introduction

In biomedical applications, the detection of ultrasound is conventionally performed with piezoelectric transducers. Nonetheless, the performance of piezoelectric technology is often inadequate in challenging applications such as intravascular photoacoustic imaging [1]-[2], where both miniaturization and high sensitivity are desired, or in magneto-acoustics [3], where immunity to electromagnetic interference (EMI) is needed. In such applications, optical interferometry represents a promising approach for ultrasound detection since it is not susceptible to EMI and can achieve miniaturization without loss of sensitivity [4]-[5].

One of the promising approaches to interferometric detection of ultrasound is the use of planar fabrication technologies to produce micron-scale optical waveguides that perform as the sensing elements. Original implementations of this approach were accomplished with polymer waveguides in which the core was made out of a polymer and the top cladding was the surrounding air or water [6]-[8]. A significant advantage of using polymers is that their photo-elastic constants are considerably larger than those of more conventional optical materials such as silica or silicon. Indeed, the reported efficiency in the conversion of acoustic to optical signals in polymer-based sensors is typically an order of magnitude higher than the one reported for silica-based sensors [9]. Nonetheless, the use of exposed waveguides limits clinical applications, in particular if the waveguide needs to come in contact with tissue. In addition, the relatively low refractive index of the polymers used has created a tradeoff between miniaturization and the Q factors achievable when such waveguides were used for ring resonators [5].

More recently, ultrasound detection has been demonstrated with waveguides fabricated in silicon-photonics technology, in particular in silicon-on-insulator (SOI) substrates [10]-[12]. The fabrication of optical waveguides in SOI substrates is performed via semiconductor fabrication techniques, characterized by their high reproducibility and mass-production capability. In addition, the high refractive index of silicon, which was used as the core material in Refs. [10]-[12], enabled a higher level of miniaturization than the one achievable by polymer cores, and the use of an over-cladding to protect the fabricated structures without significantly changing the size of the guided mode. The main disadvantage of the SOI platform is the relatively low photo-elastic coefficient of silicon, potentially limiting the sensitivity that may be achieved for ultrasound detection.

In this paper, we experimentally demonstrate that, under certain conditions, the sensitivity of SOI-based ultrasound detectors may be significantly improved by using bisbenzocyclobutene (BCB), a polymer known for its high photo-elastic coefficient [13], as the over-cladding material instead of silica. The effect of the BCB over-cladding on sensitivity was tested for both longitudinal acoustic waves and surface acoustic waves (SAWs) for TE and TM polarizations of a strip waveguide. For the case of longitudinal waves, the theoretical model developed in Ref. [12] was extended to include the effect of acoustic impedance mismatches and showed to be in good agreement with the experimental results. Our experiments show that when using the TM mode, the response of BCB-coated silicon waveguides to ultrasound (in terms of the modulation of the effective refractive index) can be comparable to the one achieved by polymer waveguides [14] and an order of magnitude higher than for the case of optical fibers. In addition, in our design, the use of a BCB over-cladding reduced the susceptibility of the detector to SAWs and acoustic reverberations – effects that can potentially limit the usability of silicon-photonics for detecting ultrasound in imaging applications.

## 2. Theoretical model

In our model, we consider the case of a strip silicon waveguide that is embedded either in a silica cladding (Fig. 1a), or in a composition of a silica under-cladding and a BCB over-cladding (Fig. 1b). For both types of waveguides, the values for the width and height of the silicon core were chosen to be 500 nm and 220 nm, respectively, in correspondence to the values offered by multi-project-wafer (MPW) services of IMEC (Leuven, Belgium) [12], whereas the total thickness of the cladding was 4 μm.

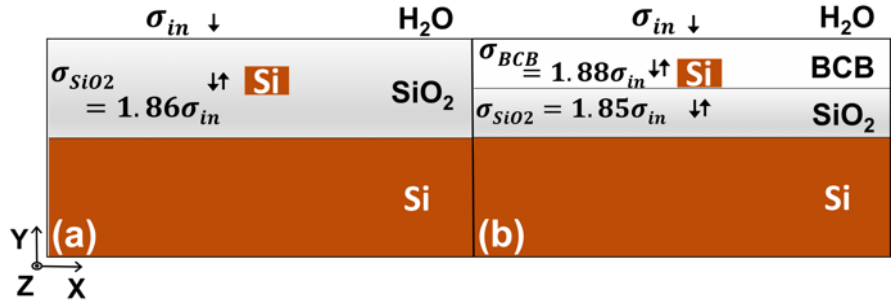

Fig. 1. The cross sections and stress σy in the layers of the two waveguides studied, which had: (a) a SiO2 over-cladding and (b) BCB over-cladding.

In our analysis, we consider longitudinal acoustic waves that impinge on the chip perpendicularly, i.e. the acoustic propagation is in the *y* direction. Denoting the acoustic impedance by W, the transmission of the stress $\sigma_y$ from medium "1" to medium "2" is given by [15]

$$t = \frac{2W_2}{W_1+W_2} \quad (1)$$

whereas the reflection is given by

$$r = \frac{W_2-W_1}{W_1+W_2} \quad (2)$$

Assuming that the silica and BCB layers in Fig. 1 are considerably small in comparison to the acoustic wavelength, the effect of multiple reflections may be modelled without accounting for the phase accumulated by the acoustic waves. Using this simplified model, one obtains that for Fig. 1a, the stress within the silica, which includes the contribution of both transmitted and reflected waves, is given by

$$\sigma_{SiO2} = \sigma_{in} \frac{2W_{Si}}{W_{H2O}+W_{Si}} \quad (3)$$

where $\sigma_{in}$ is the stress associated with the incident wave and $W_{Si}$, $W_{H2O}$ are the acoustic impedances of silicon and water, respectively. For the structure in Fig. 1b, the stresses due to the transmitted and reflected waves within the silica and BCB layers are respectively given by

$$\sigma_{SiO2} = \sigma_{in} \frac{8W_{Si}W_{BCB}(W_{H2O}W_{SiO2}+W_{BCB}^2)}{(W_{SiO2}+W_{BCB})(W_{Si}+W_{BCB})(W_{H2O}+W_{BCB})^2} \quad (4a)$$

$$\sigma_{BCB} = \sigma_{in} \frac{(W_{SiO2}+W_{Si})(W_{H2O}+W_{BCB})(W_{SiO2}^2-W_{BC}^2)-8W_{H2O}W_{SiO2}W_{BCB}(W_{Si}-W_{SiO2})}{(W_{H2O}+W_{SiO2})(W_{SiO2}+W_{Si})(W_{H2O}+W_{BCB})(W_{SiO2}+W_{BCB})} \quad (4b)$$

where here $W_{SiO2}$, $W_{BCB}$ are the acoustic impedances of silica and BCB, respectively. Using the values in Table 1, we obtain for Fig. 1a that $\sigma_{SiO2}$ = 1.86 $\sigma_{in}$ , and for Fig. 1b that $\sigma_{BCB}$ =1.88 $\sigma_{in}$ and $\sigma_{SiO2}$= 1.85 $\sigma_{in}$ . We note that the stresses $\sigma_{SiO2}$ and $\sigma_{BCB}$ in our analysis include the contribution of both the transmitted and reflected waves, as illustrated in Fig. 1.

As described in Ref. [12], there are two frequently used measures of sensitivity that describe the effect of mechanical perturbations on the optical properties of a waveguide. When one detects the change in optical phase (ϕ) in a waveguide of a given length due to uniform pressure (P), one may use the phase sensitivity: $S_\phi$ = dϕ/dP [16], [17]. When a resonator is used, the normalized sensitivity is more appropriate [9]: $S_\lambda = d\lambda_{res}/(\lambda_{res} dP)$, where $\lambda_{res}$ is the resonance wavelength and $d\lambda_{res}$ is the shift in wavelength due to the perturbation. $S_\lambda$ may be calculated by using the following equation:

$$S_\lambda = \frac{1}{n_{eff}} \frac{dn_{eff}}{dP} + \frac{d\varepsilon_z}{dP} \quad (5)$$

where $n_{eff}$ is the refractive index of guided mode and $\varepsilon_z$ is the strain in the z direction. The relation between $S_\lambda$ and $S_\phi$ is given by

$$S_\phi = \frac{2\pi n_{eff} L}{\lambda} S_\lambda \quad (6)$$

where $\lambda$ is the incident light wavelength and L is the effective length of the sensor.

To calculate $S_\lambda$ due to a plane longitudinal acoustic wave that impinges on the chip perpendicularly, we use the model of Ref. [12], in which $\varepsilon_z$, $\varepsilon_x \to 0$. The calculation of $S_\lambda$ is performed by computing the change in $n_{eff}$ of the guided mode due to deformation and change in the refractive index of the materials by the photo-elastic effect, given by the following equations:

$$\varepsilon_y = -\frac{(1+\nu)(1-2\nu)}{(1-\nu)E} \sigma_y \quad (7.a)$$

$$\Delta n_x = \frac{(C_1 \nu + C_2)}{1-\nu} \sigma_y \quad (7.b)$$

$$\Delta n_y = \frac{[(1-\nu)C_1 + 2\nu C_2]}{1-\nu} \sigma_y \quad (7.c)$$

where $C_1$ and $C_2$ are the photo-elastic constants and $\nu$ is the Poisson ratio. The values of the optical, mechanical, and acoustic parameters of silicon, silica, and BCB are summarized in Table 1. We note that for Si and SiO2, the optical parameters were measured at λ = 1550 nm [18], whereas for BCB their values were obtained at λ =1536 nm [13].

Table 1: Optical, mechanical, acoustical, and photo-elastic properties of Si, SiO2, and BCB [13], [15], [18]

| Property | Si | SiO$_2$ | BCB |
|---|---|---|---|
| Refractive index (n) | 3.48 | 1.44 | 1.54 |
| Young Modulus (E) GPa | 130 | 76.7 | 2.9 |
| Poisson Ratio (ν) | 0.27 | 0.19 | 0.34 |
| Density (ρ) Kg/m$^3$ | 2328 | 2200 | 1050 |
| Acoustic impedance (W) Kgm$^{-2}$s$^{-1}$ | 19.5 x 10$^6$ | 13.6 x 10$^6$ | 2.17 x 10$^6$ |
| Photo-elastic constant (C1) TPa$^{-1}$ | -17.13 | 1.17 | 99 |
| Photo-elastic constant (C2) TPa$^{-1}$ | 5.51 | 3.73 | 31 |

As in Ref. [12], the calculation of $S_\lambda$ via Eqs. (5) and (6) requires using a mode solver to find the perturbations to the effective refractive index. In this work, COMSOL Multiphysics was used and the analysis was conducted for the two structures shown in Fig. 1 for both the TE the TM modes. For the wavelength $\lambda = 1540$ nm, the values obtained for the TM and TE modes were $n_{eff} = 1.78$ and $n_{eff} = 2.46$, respectively, for the silica over-cladding (Fig. 1a) and $n_{eff} = 1.84$ and $n_{eff} = 2.47$, respectively, for the BCB over-cladding (Fig. 1b). Figures 2a and 2b show the electromagnetic field pattern of the TE and TM modes for the structure in Fig. 1a. We note that since the refractive index of BCB is relatively close to that of silica, the modes for the structure in Fig. 1b are similar to those shown in Fig. 2.

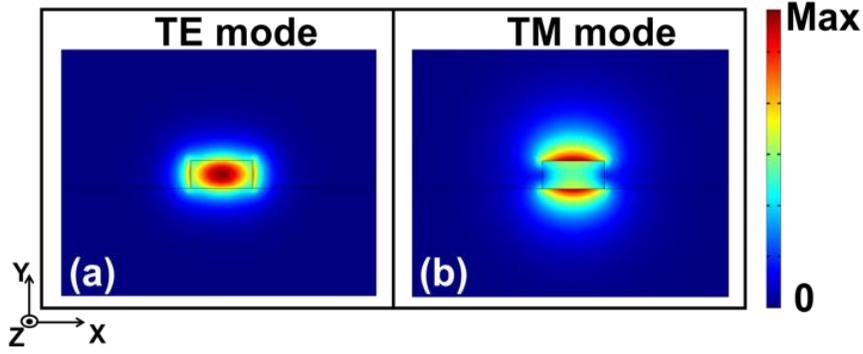

Fig. 2. The electromagnetic field of the (a) TE and (b) TM modes of the silicon waveguide shown in Fig. 1a.

## 3. Fabrication and Measurement setups

Similar to Ref. [12], the fabrication of the silicon waveguides was performed at the foundry of IMEC (Leuven, Belgium) using the SOI MWP services of ePIXfab. Two wafers were provided, in which the same silicon structures were produced. In both wafers, the silicon substrate had a thickness of 700 µm. In the first wafer, the silicon core was buried in a silica cladding, as illustrated in Fig. 1a. In the second wafer, used to produce the structure shown in Fig. 1b, the core was covered by a protective resist, which we replaced by BCB (3022-35 series, The Dow Chemical Company) using the procedure described in the following. The resist cladding was removed using acetone and the exposed wafer was spin-coated with BCB at 3000 rpm for 60 seconds, resulting in a BCB layer with a thickness of approximately 2 µm. This was followed by baking on a hotplate at 120ºC for 10 minutes to remove solvents and to stabilize the BCB film. Afterwards, the film underwent a curing process at 230ºC for 30 minutes in inert atmosphere while flowing $N_2$ gas was used to prevent oxidation.

In both wafers, 2 mm long silicon waveguides were produced with fiber-to-chip grating couplers on both ends. Polarization maintaining (PM) fibers were coupled to the waveguides using the procedure described in Ref. [12], where the orientation of the fiber with respect to the grating coupler determined whether the TE or TM mode would be launched. In total, 4 fiber-coupled chips were produced for the discussed options of polarization (TE or TM) and over-cladding material (silica or BCB).

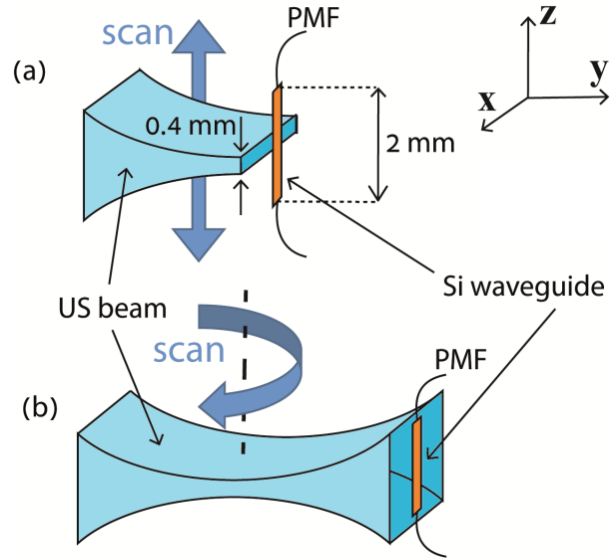

Fig. 3. Schematic diagram of the experimental setup. An ultrasound (US) beam focused in one dimension was generated using a cylindrically focused US transducer, whereas the detection was performed with the silicon strip waveguide, shown in Figs. 1a and 1b, for both the TE and TM modes, where polarization-maintaining fibers (PMFs) were used to couple it to an interferometric system that measured the US-induced phase modulation in the waveguide. The substrate of the strip waveguide is not shown in the illustration. (a) To measure $S_\lambda$, the waveguide was positioned at the focus of the transducer, at a distance of 25.76 mm, where scanning in the x and z directions was used to position the beam in the center of the waveguide. (b) To assess the effect of surface acoustic waves (SAWs) on the waveguide, it was positioned in the far field of the transducer at a distance of 49 mm, where the wavefront of the US beam was approximately planar over the length of the waveguide. Since the excitation of SAWs require an incidence angle of approximately 20 degrees, the US beam was rotated around the z axis.

We experimentally characterized the response of the silicon waveguide to ultrasound for both longitudinal acoustic waves and SAWs. In both cases, the acoustic waves were ultrasound bursts generated by a cylindrically focused ultrasound transducer with a diameter of 12.7 mm, focal length of 25.76 mm, and a central frequency of 15 MHz (Olympus, V319). The resulting phase variations in the waveguides at λ =1540 nm was monitored using the interferometric setup described in Ref. [12]. In addition, a calibrated needle hydrophone with a diameter of 40 µm, bandwidth of 30 MHz, and calibration accuracy of ±15% (Precision Acoustics) was scanned along the focus of the transducer to characterize the generated acoustic beam, yielding a maximum peak-to-peak pressure of approximately 1.3 MPa in the focus, a focal full-width-at-half-maximum (FWHM) of 0.4 mm.

The phase modulation of the light guided in the four silicon waveguides was measured in two acoustic configurations. In the first configuration, the acoustic wave was focused in the z direction on the center of the waveguide, as illustrated in Fig. 3a. The chip was scanned in the x and z directions, and the phase modulation due to the ultrasound burst was recorded for the position in which the acoustic beam was symmetrically centered on the waveguide. To compare the response of the silicon waveguide to that of the fibers, the acoustic beam was subsequently scanned away from the silicon waveguide, approximately 5 mm in the z direction, such that its focus lied entirely on the optical fibers. In the second configuration (Fig. 3b), the goal was to excite the SAW in the chips by an approximately planar acoustic wave that hits the chips at an angle of approximately 20°. Accordingly, the chip was moved to the far-field of the acoustic beam at a distance of approximately 49 mm from the transducer. The transducer was rotated around the z axis, creating an angle of θ with the normal to the chip (y axis in Fig. 3b), where the scanning in θ was performed from 0 to 30°. For each angle, the transducer was scanned in the x direction to find the position of strongest signal.

## 4. Results

Figure 4 shows the response to longitudinal acoustic waves measured using the geometry of Fig. 3a for TM (Fig. 4a) and TE (Fig. 4b) waveguides with the BCB (blue curve) and silica (red curve) over-cladding. The responses for the waveguides are compared to the signals obtained when the ultrasound beam was focused on the fibers (green dashed curve). The results in Fig. 4 show that for the TE mode, compared to SiO$_2$ over-cladding, the BCB over-cladding enhanced the signal by a factor of 1.41, whereas for the TM mode, an enhancement of 4.98 was achieved. Using Eq. (6), and the effective refractive indices of the different configurations, calculated in Section 2, the enhancement in $S_\lambda$ due to the BCB over-cladding for the TE and TM modes were 1.4 and 4.82, respectively. Using our theoretical model, and accounting for a 10 nm fabrication error in each of the dimensions of the waveguide, the theoretical values obtained for the enhancement in $S_\lambda$ were 1.13 ± 0.27 and 3.9 ± 2.3 for the TE and TM modes, respectively, in good agreement with the experimental values. As Fig. 4 shows, the signal enhancement of the BCB-coated chips with respect to the optical fibers was even higher; in terms of $S_\lambda$ the enhancement was 1.44 and 9.41 for the TE and TM modes, assuming an $n_{\text{eff}} = 1.47$ for the optical fiber in Eq. (6).

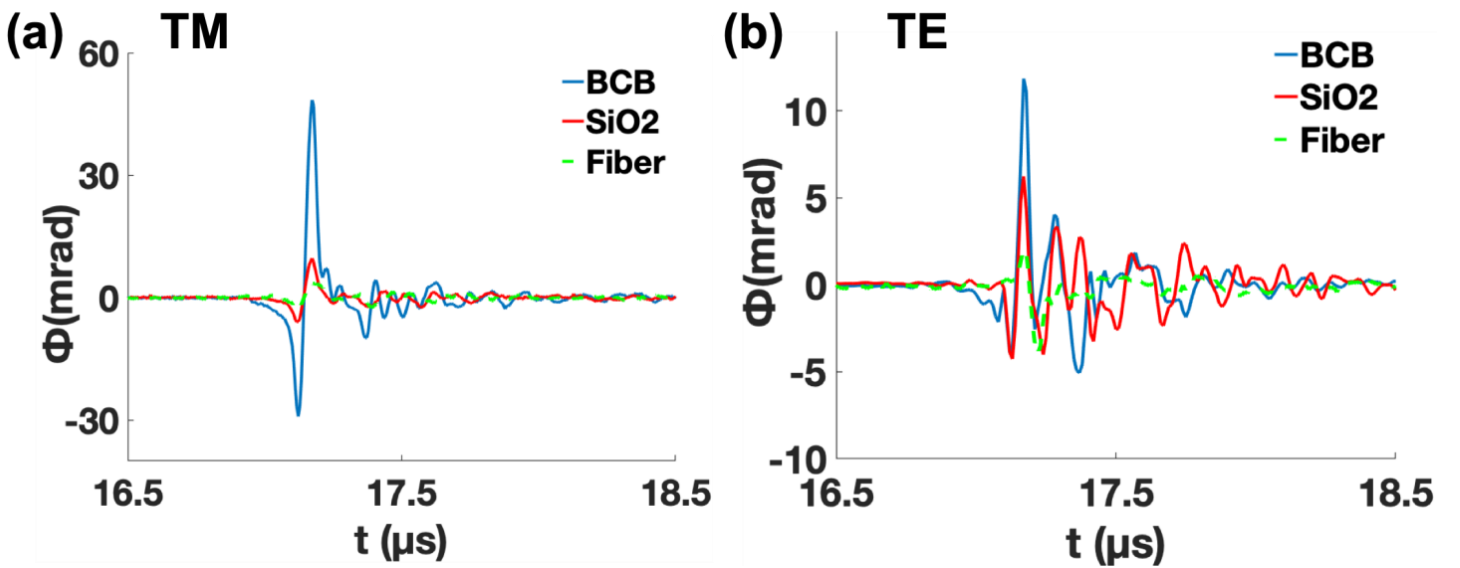

Fig. 4. Measured optical phase shifts in the silicon waveguides due to a focused ultrasound burst (Fig. 3a) and the optical phase shifts obtained from the fibers. The results are shown for the waveguides with the BCB over-cladding (blue solid curve), SiO$_2$ over-cladding (red solid curve) and from the fibers (green dashed curve) for both the (a) TM and (b) TE modes.

In Fig. 5 we normalized the responses of the waveguides and compared them to the response obtained from the hydrophone measurement. Since the dimensions of the hydrophone were different than those of the silicon waveguides, the hydrophone response was averaged over a scan length of 2 mm in the *z* direction centered on the transducer focus, effectively emulating the spatial-averaging effect experienced by the 2 mm long waveguide. As Fig. 5 shows, the initial bipolar signal was almost identical for all the waveguides and exhibited the same profile as the signal measured by the hydrophone. For both the TM and TE modes, the accompanying reverberations were reduced when the BCB over-cladding was used. Examining the peak-to-peak value of the reverberations after $t = 17.5\ \mu s$ in Fig. 5, the following values were obtained: 0.16 (TM-BCB), 0.34 (TM-SiO$_2$), 0.16 (TE-BCB), 0.5 (TE-SiO$_2$).

Since the initial bipolar signal measured with the silicon waveguides represent the average acoustic signal integrated over a length of 2 mm, the sensitivity $S_\lambda$ of the BCB-coated waveguides may be quantified using the hydrophone measurement. While the peak-to-peak pressure at the focus of the transducer was 1.3 MPa, the average signal over the 2 mm length was 0.26 kPa, leading to $S_\phi = 0.31\ \mathrm{rad\ MPa^{-1}}$ and $S_\phi = 0.055\ \mathrm{rad\ MPa^{-1}}$ for the BCB-coated TM and TE waveguides, respectively. Using Eq. 6 and accounting for the hydrophone calibration accuracy, one obtains $S_\lambda = (21 \pm 3.2) \times 10^{-6}\ \mathrm{MPa^{-1}}$ and $S_\lambda = (2.7 \pm 0.41) \times 10^{-6}\ \mathrm{MPa^{-1}}$ for the BCB-coated TM and TE waveguides, respectively.

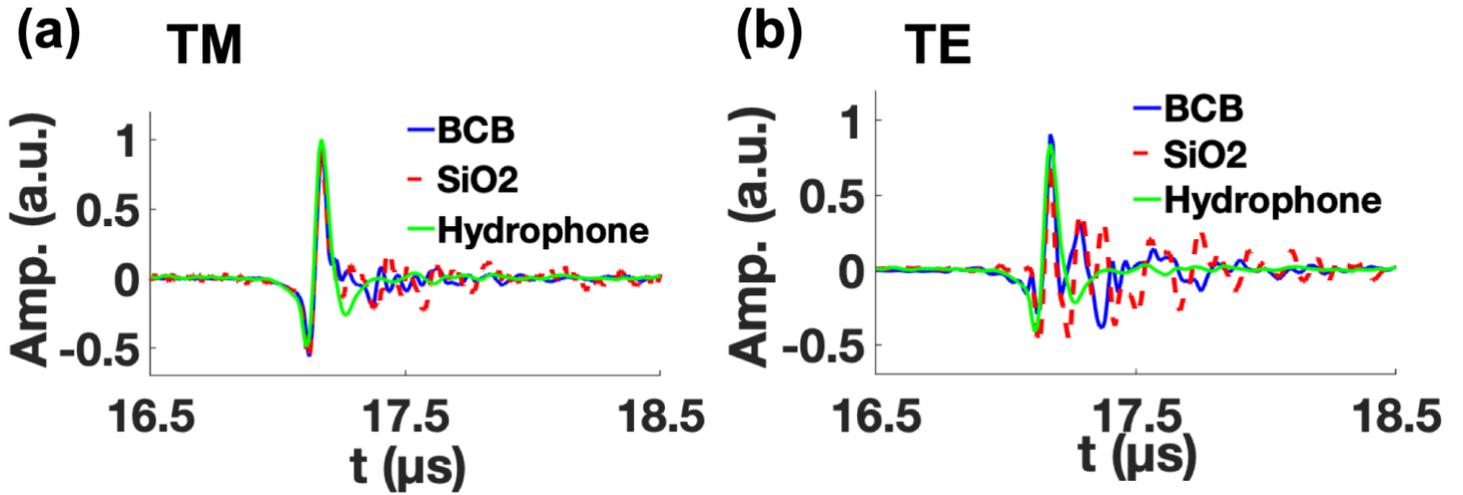

Fig. 5. Normalized response from the silicon waveguides compared to the hydrophone response due to a focused ultrasound burst. The results given are the hydrophone response (green solid curve) in comparison with the waveguides with the BCB over-cladding (blue solid curve) and SiO$_2$ over-cladding (red dashed curve) for both the (a) TM and (b) TE modes.

Figure 6 shows the peak-to-peak phase values obtained as a function of angle in the second acoustic configuration in which the response to SAWs was assessed (Fig. 3b). The results are shown for BCB- (blue curve) and silica-coated (red curve) chips for the TM (Fig. 6a) and TE (Fig. 6b) waveguides. The effect of SAWs is clearly visible in Fig. 6 as the large variations in the signal around the angle of 20°. As expected from the results of Refs. [9] and [12], SAWs dominate the response of the silica-cladding chips for the TE mode, whereas the response for the TM mode is more moderate. For the chips with the BCB over-cladding, the response of the TM mode to SAWs was comparable in magnitude to that of the silica-cladding chips, whereas in the TE waveguide the effect of SAWs was diminished by over an order of magnitude. We note that while the responses in Fig. 6, which were measured in the far field, also include the contribution of phase perturbation in the fibers in the proximity of the silicon waveguide, the clear qualitative and quantitative differences between the responses may be solely attributed to susceptibility of the silicon waveguides in the 4 chips to SAWs.

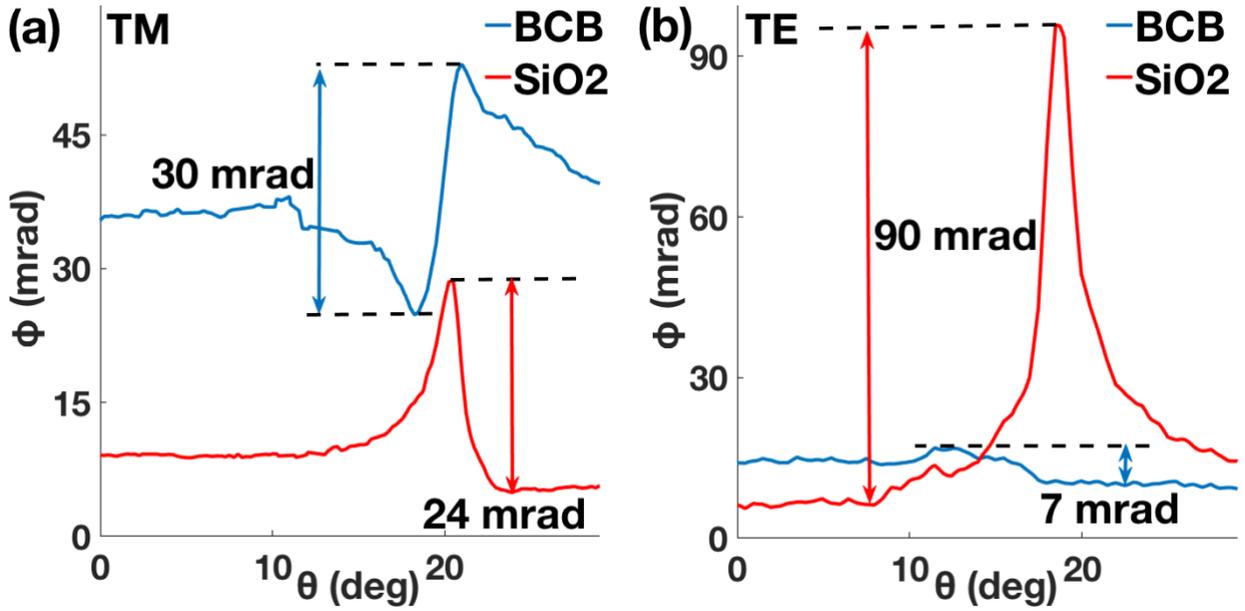

Fig. 6. Measured peak-to-peak values of the ultrasound-induced phase shifts in the far-field measurement illustrated in Fig. 3b obtained for the waveguides with BCB over-cladding (blue curve) and SiO$_2$ over-cladding (red curve) for (a) TM and (b) TE modes.

## 4. Conclusion

In conclusion, we experimentally demonstrated that a BCB over-cladding can significantly improve the capabilities of silicon-photonics waveguide for ultrasound detection in terms of sensitivity and susceptibility to acoustic reverberations and SAWs. For longitudinal waves, the enhancement in S$_\lambda$ was approximately 4.82 for the TM mode and 1.4 for the TE mode, in agreement with our theoretical model. In our analysis, we introduced a modification to the theoretical model of Ref. [12] that takes into account the effect of reflection and refraction in the cladding due to acoustic impedance mismatches. According to our model, the effect of BCB on the propagation of the acoustic waves is negligible (Fig. 1b) and the enhancement in S$_\lambda$ may be solely attributed to the high photo-elastic coefficients of BCB. The higher enhancement for the TM mode may be explained by its spatial distribution (Fig. 2b), whose overlap with the over-cladding is more significant than in the case of the TE mode. Although our analysis is limited to the case of an acoustic wave with normal incidence, the angular dependence of the acoustic response was experimentally studied to evaluate the effect of SAWs. For SAWs, the use of a BCB over-cladding led to an order of magnitude reduction in the response of the TE mode, whereas for the TM the magnitude of the response did not change considerably. Nonetheless, because of its significant enhancement in sensitivity to longitudinal waves, the relative susceptibility of the TM mode to SAWs was also significantly reduced by the BCB coating. We note that a full analysis of the response to SAWs for BCB-coated chips is considerably more complex than the one performed in Ref. [12] owing to the multi-layer structure of such chips.

In all the measurements, the initial bipolar signal detected by the silicon waveguides was accompanied by acoustic reverberations, which may be explained by the impedance mismatch between the silicon substrate and its surrounding. Since the speed of sound in silicon is approximately 8400 m/s, the acoustic roundtrip in the 700 μm thick silicon substrate was 0.17 μs, comparable to the duration of the ultrasound burst, as measured by the hydrophone. Thus, the reverberations due to the thickness of the substrate are expected to continually appear on the measured signals, rather than appear at distinct time instances. In both the TE and TM waveguides, the use of BCB for the over-cladding reduced the effect of reverberations, where the weakest effect was obtained for the BCB-coated TM waveguides. Further reduction in the effect of the reverberation may be achieved by using an additional lossy backing layer with high acoustic impedance, similar to those used in piezoelectric transducers [19], or by locally thinning of the silicon substrate below the waveguides to a thickness that is considerably smaller than the acoustic wavelength.

The use of a silicon waveguide with a polymer over-cladding may be regarded as a hybrid approach that exploits the advantages of both materials. Using this approach, optical resonators may be produced that achieve the miniaturization level offered by silicon photonics and combine it with S$_\lambda$ values comparable to those achieved with polymer waveguides. For example, in Ref. [14] a sensitivity of $S_\lambda = 21 \times 10^{-6}$ MPa$^{-1}$ was measured for a polystyrene microring, which is equal to the value achieved in this work for the BCB-coated TM waveguide. Further enhancement in S$_\lambda$, beyond the one reported in this work may be achieved by exposing more of the guided mode to the polymer over-cladding.

## Funding

Israel Science Foundation (ISF) (694/15, 942/15); Volkswagen Foundation (ZN3172); EVPR fund at the Technion.

## Acknowledgements

The authors would like to thank the Technion Micro–Nano Fabrication and Printing Unit (MNF&PU) for the technical support with the sample preparation.

## References

[1] S. Sethuraman, J. H. Amirian, S. H. Litovsky, R. W. Smalling, and S. Y. Emelianov, "Spectroscopic intravascular photoacoustic imaging to differentiate atherosclerotic plaques," Opt Express, vol. 16, no. 5, pp. 3362–3367, 2008.
[2] K. Jansen, A. F. W. van der Steen, H. M. M. van Beusekom, J. W. Oosterhuis, and G. van Soest, "Intravascular photoacoustic imaging of human coronary atherosclerosis," Optics Letters, vol. 36, no. 5, pp. 597-599, 2011.
[3] S. Kellnberger, A. Rosenthal, A. Myklatun, G. G. Westmeyer, G. Sergiadis, and V. Ntziachristos, "Magnetoacoustic Sensing of Magnetic Nanoparticles," Physical Review Letters, vol. 116, no. 10, pp. 108103, 2016.
[4] G. Wissmeyer, M. Pleitez, A. Rosenthal, and V. Ntziachristos, "Looking at sound: optoacoustics with all-optical ultrasound detection," Light: Science and Applications, vol. 7, no. 1, pp. 53, 2018
[5] C. Zhang, S. Chen, T. Ling and L. Guo, "Review of imprinted polymer microrings as ultrasound detectors: Design, fabrication, and characterization", IEEE Sensors Journal, vol. 15, no. 6, pp. 3241–3248, 2015.
[6] C. Chao, L. Guo, S. Ashkenazi and M. O'Donnell, "Ultrasound detection using polymer microring resonators," in Conference on Lasers and Electro-Optics (CLEO), 2005, vol. 1, pp. 758-760.
[7] S. Chen, S. Huang, T. Ling, T, S. Ashkenazi and L. Guo, "Polymer microring resonators for high-sensitivity and wideband photoacoustic imaging," IEEE Transactions on Ultrasonics, Ferroelectrics, and Frequency Control, vol. 56, no. 11, pp. 2482–2491, 2009.
[8] V. Govindan and S. Ashkenazi, "Bragg waveguide ultrasound detectors," IEEE Transactions on Ultrasonics, Ferroelectrics, and Frequency Control, vol. 59 no. 10, pp. 2304–2311, 2012.
[9] A. Rosenthal, S. Kellnberger, D. Bozhko, A. Chekkoury, M. Omar, D. Razansky, and V. Ntziachristos, "Sensitive interferometric detection of ultrasound for minimally invasive clinical imaging applications," Laser and Photonics Reviews, vol. 8, no. 3, pp. 450–457, 2014.
[10] A. Rosenthal, M. Omar, H. Estrada, S. Kellnberger, D. Razansky., and V. Ntziachristos, "Embedded ultrasound sensor in a silicon-on-insulator photonic platform," Applied Physics Letters, vol. 104, no. 2, pp. 021116, 2014.
[11] S. M. Leinders, W. J. Westerveld, J. Pozo, P. L. M. J. Van Neer, B. Snyder, P. O'Brien, H. P. Urbach, N.de Jong and M. D. Verweij, "A sensitive optical micro-machined ultrasound sensor (OMUS) based on a silicon photonic ring resonator on an acoustical membrane," Scientific Reports, vol. 5, no. 1, pp. 14328, 2015.
[12] S. Tsesses, D. Aronovich, A. Grinberg, E. Hahamovich, and A. Rosenthal, "Modeling the sensitivity dependence of silicon-photonics-based ultrasound detectors," Optics Letters, vol. 42, no. 24, pp. 5262, 2017.
[13] M. F. Hossain, H. P. Chan and M. A. Uddin, "Simultaneous measurement of thermo-optic and stress-optic coefficients of polymer thin films using prism coupler technique", Appl. Opt., vol. 49, no. 3, pp. 403–408, 2010.
[14] C. Chao, S. Ashkenazi, S. Huang, M. O'Donnell and L. Guo, "High-Frequency Ultrasound Sensors Using polymer microring resonators," IEEE Transactions on Ultrasonics, Ferroelectrics, and Frequency Control, vol. 54, no. 5, pp. 957–965, 2007.
[15] J. L. Rose, "Ultrasonic guided waves in solid media," Cambridge University Press, 2014.
[16] E. Zhang, J. Laufer and P. Beard, "Backward-mode multiwavelength photoacoustic scanner using a planar Fabry-Perot polymer film ultrasound sensor for high-resolution three-dimensional imaging of biological tissues," Applied Optics, vol. 47, no. 4, pp. 561,2008.
[17] G. N. De Brabander, J. T. Boyd and G. Beheim, "Integrated Optical Ring Resonator with Micromechanical Diaphragm for Pressure Sensing," IEEE Photonics Technology Letters, vol. 6, no. 5, pp. 671–673, 1994.
[18] D. J. Lockwood and L. Pavesi, "Silicon photonics II: components and integration," Springer, 2011.
[19] J. M. Cannata, T. A. Ritter, W. H. Chen, R. H. Silverman, and K. K. Shung, "Design of efficient, broadband single-element (20-80 MHz) ultrasonic transducers for medical imaging applications," IEEE Transactions on Ultrasonics, Ferroelectrics, and Frequency Control, vol. 50, no. 11, pp. 1548–155, 2003.